\documentclass[fleqn,twoside]{article}
\usepackage{espcrc2}

\usepackage{graphicx}
\usepackage[figuresright]{rotating}

\newcommand{\AmS}{{\protect\the\textfont2
  A\kern-.1667em\lower.5ex\hbox{M}\kern-.125emS}}

\title{Radiative and nonleptonic hyperon decays in broken SU(3)}

\author{P. \.Zenczykowski\address{Division of Theoretical Physics, Institute of Nuclear Physics, Polish Academy
of Sciences, Radzikowskiego 152, 31-342 Krak\'ow, Poland}%
        \thanks{This work was supported in part by the Polish State Committee
	for Scientific Research Grant No. 2 P03B 046 25.}}
       
\begin{document}

\begin{abstract}
We report on the recently proposed joint resolution 
of two long-standing puzzles 
in weak radiative (WR) and nonleptonic (NL) hyperon decays (HD). 
First, a good VMD-based description of WRHD is presented. In particular,
the large negative asymmetry observed in the $\Sigma^+ \to p\gamma$ decay
is explained through a calculably large SU(3)-breaking effect 
in the relevant parity-violating amplitude.
Second,  
the achieved description of the parity-violating WRHD amplitudes
permits the extraction, via the $SU(2)_W$ + VMD route,
of the non-soft-pion correction term in the
parity-violating NLHD amplitudes.
The latter subtracts a substantial amount from the
current-algebra commutator term, thus leading to the resolution of the old $S:P$
discrepancy in NLHD.
\vspace{1pc}
\end{abstract}

\maketitle

\section{THE PUZZLES}

For a long time weak hyperon decays have presented us with two puzzles: 
the problem of
the $S:P$ ratio in nonleptonic decays and the issue of a large negative
asymmetry in the $\Sigma^+\to p\gamma$ radiative decay.

\subsection{{The $S:P$ problem}}

The $S:P$ puzzle is about 50 years old. 
It concerns the relative size of parity-violating (p.v., $S$-wave) 
and parity-conserving (p.c., $P$-wave) amplitudes in nonleptonic hyperon decays such
as $\Sigma^+\to p\pi^0$ etc.
Each of the two sets of amplitudes may be described in terms of two
$SU(3)$ amplitudes: $f$ and $d$. 
Their experimental values, as determined in
\cite{DGH1986}, are (in units of $10^{-5} MeV$) -\\ 
\phantom{}in the p.c. sector:
\begin{eqnarray}
\label{fPdP}
f_P=4.7, &
\phantom{andx} &d_P=-2.6, 
\end{eqnarray}
\phantom{}and in the p.v. sector:
\begin{eqnarray}
\label{fSdS}
f_S=3.0,& \phantom{andx} &d_S=-1.2.
\end{eqnarray}
Consequently, we have:
\begin{eqnarray}
\label{eq1}
d_P/d_S\approx 2.2&f_P/d_P\approx -1.8&f_S/d_S\approx -2.5
\end{eqnarray}
This constitutes {\it{Puzzle \#1}}, since the soft-pion theorems
\cite{softpiontheorems} predict that
$f_S=f_P$ and $d_S=d_P$.

\subsection{Large negative asymmetry in $\Sigma^+ \to p\gamma$}

Th $\Sigma^+ \to p\gamma$ puzzle dates back to the 1960's. 
It is composed of several ingredients:\\[3pt]
\phantom{xx}(A) Hara's theorem \cite{Hara} states that 
"Parity-violating $\Sigma^+ \to
p\gamma$  amplitude must vanish in the $SU(3)$ limit". 
Thus, with $SU(3)$-breaking effects expected to contribute
at the 20\% level,
one anticipates small asymmetry $|\alpha(\Sigma^+ \to p \gamma)|\approx 0.2$.\\ [1pt]
\phantom{xx}(B) This expectation is in disagreement with  experiment: 
$\alpha_{exp}(\Sigma^+ \to p \gamma)=-0.76\pm 0.08$.\\[1pt]
\phantom{xx}(C) In addition, there are several 
conflicts between the three main types of theoretical approaches to the
p.v. WRHD amplitudes:
\begin{enumerate}
\item The hadron-level pole model (with $1/2^-$ baryons in the intermediate state) 
\cite{Gavela1981} agrees with Hara's theorem in the $SU(3)$ limit, while for
broken $SU(3)$ it yields $\alpha(\Sigma^+ \to p \gamma)\approx -0.8$, despite
expectations in (A).
\vspace{-4pt}
\item The simple quark model \cite{KR1983} violates Hara's theorem in the
$SU(3)$
limit, and yields $\alpha(\Sigma^+ \to p \gamma)\approx -0.6$ for broken $SU(3)$.
\vspace{-4pt}
\item The hadron-level VMD+$SU(6)$ model \cite{PZ1989} violates Hara's theorem in the
$SU(3)$ limit, and yields $\alpha(\Sigma^+ \to p \gamma)\approx -0.9$ for broken
$SU(3)$.
\end{enumerate}
The issue of the violation of Hara's theorem and the above 
theoretical conflicts constitute {\it{Puzzle \#2}}.

\section{CRUCIAL EXPERIMENTAL INPUT}

In ref.\cite{LZ1995} it was observed that the status of Hara's theorem may be
clarified through a measurement of the $\alpha(\Xi^0 \to \Lambda \gamma)$
asymmetry.   
The analyses of \cite{LZ1995} show that,
independently of whether one adopts an approach in which Hara's theorem 
is satisfied or violated in the SU(3)
limit,
the absolute value of this asymmetry 
must be large (around $0.8$), and the 
theoretical error of this estimate must be small (around $\pm 0.15$) even for strong SU(3)
breaking effects. As shown in Table 1,
the sign of this asymmetry indicates then what happens in the $SU(3)$ limit.
Theoretical errors of other asymmetries (whose signs are gathered in Table 1)
 are larger 
than in the $\Xi^0 \to \Lambda \gamma$ case, and depend on model details.
In particular, the $\Lambda \to n \gamma$ decay cannot provide reliable
information permitting to discriminate
between the models:
the relevant theoretical amplitudes involve substantial cancellations,
thus being prone to sizable errors, while the experimental determination
of the corresponding asymmetry is extremely difficult.
Theoretical estimates  
also show that the $\Xi^0 \to \Sigma^0 \gamma$ asymmetry
should be negative and fairly large (between $-0.5$ to $-1.0$), 
independently of the status of Hara's theorem.\\[2pt]

\noindent
Table 1\\
Theoretical signs of asymmetries in WRHD for broken SU(3)\\
\begin{tabular}{ccc}
\hline
Hara's theorem &&\\
in the SU(3) limit:
&satisfied & violated
\\
\hline
$\Sigma ^+ \to p \gamma$&
$-$&$-$\\
$\Lambda \to n\gamma$&$-$&$+$\\
$ \Xi ^0 \to \Lambda \gamma$&$
-0.8\pm0.15$& $+0.8\pm0.15$\\
$\Xi ^0 \to \Sigma^0 \gamma$&$-$&$-$
\end{tabular}\\[2pt]

Recent NA48 measurements \cite{NA48} have yielded: 
\begin{equation}
\alpha (\Xi^0 \to \Lambda \gamma) = -0.78 \pm 0.19, 
\end{equation}
thus proving that Hara's theorem is true in the SU(3) limit.
This leaves the following two theoretical questions to be answered:\\ 
\phantom{xx}1) what is wrong with the theoretical calculations predicting the
violation of Hara's theorem,\\
\phantom{xx}2) how (moderate) SU(3) breaking can result in the large 
$\Sigma^+ \to p \gamma$ asymmetry.

\section{WRHD: THEORETICAL RESOLUTION}

It appears that the origins of the violation of Hara's theorem in the simple quark model
\cite{KR1983} and in the VMD+$SU(6)$ approach \cite{PZ1989} are different.

Quark model calculations of ref.\cite{KR1983} 
and its sequels violate confinement: in the $SU(3)$ limit,
the quark propagating between the actions of the weak and electromagnetic
Hamiltonians enters its mass shell and, consequently, it propagates to
infinity, away from the other two quarks \cite{PZ2001}. 
Thus, the violation of Hara's theorem in the simple quark model is an
artefact of the approach.

It is more difficult to identify 
the origin of the violation of Hara's theorem in the VMD+$SU(6)$ approach of
\cite{PZ1989}. 
Indeed, as the latter deals with hadrons only, no violation of
confinement may be even contemplated here. However,
any VMD+$SU(6)$ approach is based on a connection between the weak couplings of
pseudoscalar and vector mesons to baryons, i.e. on an understanding of how
the nonleptonic hyperon decays are linked to weak couplings of vector mesons
to baryons.
A detailed analysis \cite{PZ2003} shows 
that the particular form of the connection
between the relevant p.v. weak couplings, established in
\cite{DDH} for the needs of nuclear parity violation 
(mainly for the evaluation of the weak p.v.
$\rho NN$ couplings)
and used both in \cite{PZ1989} and in other papers, is erroneous,
as briefly explained below.

\subsection{Soft-pion approximation in NLHD}

In general, the p.v. NLHD amplitudes may be decomposed
into a sum of two terms, the current algebra commutator term and the
correction term vanishing in the soft-pion limit.
Under the assumption than in the real world the emitted pion is still
sufficiently soft so that the correction term may be neglected, the 
p.v. NLHD amplitudes are given by the commutator term only.
This is the assumption used in \cite{DDH} when evaluating the weak couplings of
pseudoscalar mesons to baryons. The corresponding
weak couplings of vector mesons to baryons are then evaluated by spin symmetry.
When VMD is added, violation of Hara's theorem and the related prediction of
 $\alpha (\Xi^0 \to \Lambda \gamma) \approx +0.8$ ensues.

\subsection{Non-soft-pion correction term}

A thorough analysis of how pions and vector mesons couple to other hadrons
in the presence of weak interactions shows 
that the commutator term in the p.v. NLHD amplitudes
 has no counterpart in the p.v. weak couplings of vector mesons to baryons
 (see e.g.\cite{softpiontheorems},\cite{PZ2003}). 
 This is so because the connection between the
pseudoscalar meson field and the axial current is different from the connection
between the vector meson field and the vector current.
As a result, any non-zero weak p.v. coupling of vector mesons to baryons 
must be related via spin symmetry to the non-soft-pion correction term
in the p.v. NLHD amplitudes \cite{PZ2003}.
Symmetry properties of the non-soft-pion term determine then
the corresponding properties 
of its spin-symmetry-related vector-meson counterpart.
When VMD is added, these properties ensure both that  
Hara's theorem is satisfied and that 
$\alpha (\Xi^0 \to \Lambda \gamma) \approx -0.8$.

\section{WRHD IN BROKEN $SU(3)$}

The qualitative explanation given above was recently backed up by a 
model calculation where $SU(3)$ symmetry breaking was explicitly taken into
account \cite{PZ2006}.

Ref.\cite{PZ2006} uses the $SU(3)$-breaking pole model for the
p.c. NLHD amplitudes, and extracts
the P-wave parameters $f_P$ and $d_P$ from the data. The extracted values are
$f_P=5.8\times10^{-5}$ and $d_P=-3\times 10^{-5}$ 
(comparison with Eq.(\ref{fPdP}) indicates the size of expected errors).
Application of spin symmetry and VMD leads to parameter-free
predictions for the
p.c. WRHD amplitudes.

Data on the WRHD branching ratios and asymmetries, when supplied with the
above-determined
p.c. WRHD amplitudes and the SU(3)-breaking 
pole model for the p.v. WRHD amplitudes, permit the extraction
of the parameters describing the latter amplitudes.
The relevant parameters (defined in \cite{PZ2006}, see also Section
\ref{SectionNLHD}), corresponding to two-quark ($b_R$)
and single-quark ($s_R$)
contributions, are (in units of $10^{-7}$):
\begin{eqnarray}
b_R\approx 5.3,&\phantom{andx}&s_R\approx -0.75
\end{eqnarray}
Description of the WRHD data achieved with these parameters 
is given in Tables 2 and 3.

When judging the description of WRHD given in Tables 2 and 3, one has to
consider the facts that the branching ratios of the $\Sigma^+ \to p\gamma$
and $\Lambda \to n \gamma$ decays are sensitive both
to the details of the $SU(3)$ breaking
effects and to various cancellations, and that the observed discrepancies 
correspond to errors of the order of 20\% at the amplitude level.
Analyses \cite{LZ1995} show that the comparison of data and the model 
is most reliable for the $\Xi^0 \to \Lambda \gamma$
decay.\\[-2pt]

\noindent
Table 2\\
WRHD branching ratios \\
\begin{tabular}{ccc}
\hline
&data (\cite{PDG},\cite{NA48})& model (\cite{PZ2006})
\\
\hline
$\Sigma ^+ \to p \gamma$      &  $1.23\pm0.05$  &   $0.72$\\
$\Lambda \to n\gamma$         &  $1.75\pm0.15$  &   $0.77$\\
$ \Xi ^0 \to \Lambda \gamma$  &  $1.16\pm0.08$  &   $1.02$\\
$\Xi ^0 \to \Sigma^0 \gamma$  &  $3.33\pm0.10$  &   $4.42$\\
$\Xi^- \to \Sigma^- \gamma $  &  $0.127\pm0.023$&   $0.16$
\end{tabular}\\[2pt]

\noindent
Table 3\\
WRHD asymmetries \\
\begin{tabular}{ccc}
\hline
&data (\cite{PDG},\cite{NA48}) & model (\cite{PZ2006})
\\
\hline
$\Sigma ^+ \to p \gamma$      &  $-0.76\pm0.08$  &   $-0.67$\\
$\Lambda \to n\gamma$         &                  &   $-0.93$\\
$ \Xi ^0 \to \Lambda \gamma$  &  $-0.78\pm0.19$  &   $-0.97$\\
$\Xi ^0 \to \Sigma^0 \gamma$  &  $-0.63\pm0.09$  &   $-0.92$\\
$\Xi^- \to \Sigma^- \gamma $  &  $+1.0\pm1.3$    &   $+0.80$
\end{tabular}\\[2pt]

The size of the $SU(3)$ breaking effects is controlled by the dependence 
of the  p.v. WRHD amplitudes on the value of the $SU(3)$-breaking parameter
$x\equiv \Delta m_s/\Delta \omega \approx
1/3$, where $\Delta m_s \approx 190~MeV $ is an
estimate of the strange-nonstrange quark mass difference, and $\Delta \omega \approx
570~MeV$ is an estimate of the mass difference between the $1/2^+$ and $1/2^-$
baryons.

Table 4 shows relative sizes of the dominant two-quark contributions 
in the $SU(3)$ symmetric ($x=0$) and $SU(3)$ breaking
($x=1/3$) cases.
Note that while the $SU(3)$ breaking effects for the relevant $\Xi^0$ amplitudes
are of the order of 30\%, the corresponding change in the size of the $\Sigma^+ \to p\gamma$ amplitude is
enormous. Indeed, the $SU(3)$-breaking contribution to the latter amplitude 
is of the same order of magnitude as the
$SU(3)$-symmetric contributions in the $\Lambda \to n\gamma$, $\Xi^0\to \Lambda
\gamma$ and $\Xi^0 \to \Sigma^0\gamma$ amplitudes.
This completes the resolution of {\it Puzzle \#2}.

\noindent
Table 4\\
Relative size of two-quark contribution to p.v. WRHD amplitudes (\cite{PZ2006})\\
\begin{tabular}{ccc}
\hline
&SU(3) exact & SU(3) broken
\\
\hline
$\Sigma ^+ \to p \gamma$      &  $0$       &   $+0.196$\\
$\Lambda \to n\gamma$         &  $+0.192$   &   $+0.048$\\
$ \Xi ^0 \to \Lambda \gamma$  &  $-0.192$  &   $-0.128$\\
$\Xi ^0 \to \Sigma^0 \gamma$  &  $-0.333$  &   $-0.5$
\end{tabular}

\section{NON-SOFT-PION TERMS IN NLHD}
\label{SectionNLHD}
When VMD and spin symmetry are taken into account, 
parameters $b_R$ and $s_R$ describing the p.v. WRHD amplitudes
are related to the non-soft-pion
correction terms in NLHD. 
For $SU(3)$ symmetric denominators in the $1/2^-$ pole model, 
the modified connection between the $S$- and $P$ -wave amplitudes is: 
$f_S=f_P+f_R$, $d_S=d_P+d_R$, where
amplitudes
$f_{S(P)}$, $d_{S(P)}$ (corrections $f_R$, $d_R$)
are related to
the two-quark $b_{S(P)}$
and single-quark $c_{S(P)}$
($b_R$, $c_R$) amplitudes through \cite{PZ2006}
\begin{eqnarray}
b=4d/F_{\pi},&\phantom{and}&c=6(f+d)/F_{\pi}
\end{eqnarray}
with $F_{\pi}=94~MeV$.

In terms of the two-quark and single-quark amplitudes $b$ and $c$,
the soft-pion prediction $f_S=f_P$, $d_S=d_P$ then reads
\begin{eqnarray}
\label{eqbc}
b_S=b_P,&\phantom{andxx}& 
c_S=c_P. 
\end{eqnarray}
For the values of $f,d$ from
Eqs (\ref{fPdP},\ref{fSdS}), the above equations read (in units of $10^{-7}$)
\begin{eqnarray}
-5 \approx -11.1,&\phantom{xx}& 12\approx 13.4.
\end{eqnarray}
While $c_R$ is not identical to $s_R$, 
the smallness of single-quark ($s_R$) effects in
WRHD suggests the neglect of
$c_R$ \cite{PZ2006}.
For broken $SU(3)$ and with the non-soft-pion term,
Eq.(\ref{eqbc}) is then modified to 
\begin{eqnarray}
b_S=b_P+b_R/(1+x)&&c_S\approx c_P
\end{eqnarray}
which for $x = 1/3$ reads
\begin{eqnarray}
-5 \approx -6.6, &\phantom{and}& 12 \approx 13.4.
\end{eqnarray}
For the $SU(3)$ case with $x=0$, one obtains 
\begin{eqnarray}
\label{eqfin}
d_P/d_S\approx 2 &f_P/d_P\approx -1.8&f_S/d_S\approx -2.6
\end{eqnarray}
in excellent agreement with Eq.(\ref{eq1}).
Thus, the differences between the values of $f$ and $d$ as 
evaluated from $S$- and $P$- wave amplitudes result mainly
from the two-quark contribution to the non-soft-pion correction term.
In this way, {\it Puzzle \#1} is related to the fact that
the WRHD amplitudes are dominated by the two-quark terms.
\section{CONCLUSIONS}
Simultaneous explanation of the size of $SU(3)$ breaking in the
$\Sigma^+ \to p \gamma$ p.v. amplitude and of the difference between the values
of the $S$- and $P$- wave $SU(3)$ parameters in the NLHD amplitudes 
indicates that weak hyperon decays have been finally understood.
However, this understanding indicates at the same time that 
there might be a problem in
the description of nuclear
parity violation as the standard approach \cite{DDH} uses
an erroneous connection between the weak couplings of pseudoscalar and vector
mesons to baryons.

\end{document}